\documentclass{aip-cp}

\usepackage[numbers]{natbib}
\usepackage{rotating}
\usepackage{graphicx}

\begin{document}

\title{EBL Constraints Using a Sample of TeV Gamma-Ray Emitters Measured with the MAGIC Telescopes}

\author[aff1,aff2]{D. Mazin\corref{cor1}}
\author[aff3]{A.~Dom\'inguez}
\author[aff4]{V.~Fallah Ramazani}
\author[aff3]{T.~Hassan}
\author[aff5]{A.~Moralejo}
\author[aff3]{M.~Nievas Rosillo}
\author[aff6,aff7]{G.~Vanzo}
\author[aff6,aff7]{and M.~V\'azquez Acosta}
\author[]{for the MAGIC Collaboration}
\eaddress{wwwmagic.mpp.mpg.de}

\affil[aff1]{Max-Planck-Institut f\"ur Physik, D-80805 M\"unchen, Germany.}
\affil[aff2]{Institute for Cosmic Ray Research, University of Tokyo, 277-8582 Chiba, Japan.}
\affil[aff3]{Universidad Complutense, E-28040 Madrid, Spain.}
\affil[aff4]{Tuorla observatory, University of Turku, 21500 Piikki\"o, Finland.}
\affil[aff5]{Institut de Fisica d'Altes Energies, The Barcelona Institute of Science and Technology, 08193 Bellaterra, Spain.}
\affil[aff6]{Institut de Astrof\'isica de Canarias, E-38200 La Laguna, Tenerife, Spain.}
\affil[aff7]{Universidad de La Laguna, Dpto. Astrof\'isica, E-38206 La Laguna, Tenerife, Spain.}
\corresp[cor1]{Corresponding author: mazin@mpp.mpg.de}

\maketitle

\begin{abstract}
MAGIC is a stereoscopic system of two Imaging Atmospheric Cherenkov Telescopes
operating in the very high energy (VHE) range from about 50 GeV to over 50 TeV.
The VHE gamma-ray spectra measured at Earth carry an imprint of the
extragalactic background light (EBL) and can be used to study the EBL density
and its evolution in time. In the last few years, precision measurements of
several blazars in the redshift range from z=0.03 up to z=0.9 were performed
with MAGIC obtaining strong limits on the EBL density from single sources. In
this paper, we present the results from a combined likelihood analysis using
this broad redshift range sample of blazars allowing us to probe the EBL at
different wavelengths. The implications on the EBL models and perspectives for
future observations with MAGIC are also discussed.  
\end{abstract}

\section{INTRODUCTION}
The observation of bright extragalactic sources at
very-high-energy (VHE, E$>$100GeV) $\gamma$ rays
allows one to study the low energy photon background.
This background radiation, called the Extragalactic Background Light
causes energy and redshift dependent absorption of the VHE fluxes \citep{Gould1967}.
Thus, the observed spectra of extragalactic sources carry an imprint of the 
EBL density, which can be inferred from the VHE observations 
\citep[see, e.g.,][for reviews]{HauserDwek2001,DwekKrennrich2013}. 
Recent studies used 
Fermi-LAT and H.E.S.S.\ data to draw strong constraints and even detection of the 
EBL in UV, optical, and near infrared regimes \citep{Fermi2012, HESS2013}. 
Also several studies have been carried out to derive EBL upper and lower limits using published energy spectra
of extragalactic VHE $\gamma$-ray sources
\citep[see, e.g.,][]{Mazin2007,MeyerEBLupdate2012,DominguezHorizon2013,Biteau2015ApJ}.

The EBL models have been improving in the last years, too, thanks to the rich data available on
galaxies at redshifts up to z$\sim1$ with instruments like HST, ISO and Spitzer. The recent models
of \cite{Franceschini2008,Finke2010,Dominguez2011,Gilmore2012} can reproduce well the EBL spectrum seen
at redshift z=0 and have in general good agreement among each other as well as with the indirect constraints
and detections using VHE $\gamma$-ray sources. 
The latter suggests that there are not many unresolved sources of EBL out there.
Still, no convincing wavelength-resolved and model independent EBL detection could be achieved so far, 
and EBL evolution, in particular above z$>$0.5, is not resolved in direct or indirect measurements yet. 

The MAGIC Collaboration operating two 17\,m diameter Imaging Atmospheric Cherenkov Telescopes (IACT)
in La Palma \citep{MAGICperformance20142,MAGICperformance2014} observed a strong flare 
from high-frequency peaked BL Lac 1ES 1011+496 (z=0.212) in 2014. The data were analyzed
in a similar method using a likelihood ratio test as in \cite{Fermi2012, HESS2013}
and strong EBL limits (the strongest from a single VHE $\gamma$-ray source) 
were derived \citep{MAGIC-EBL-1011}. 

Here we use a similar method as \cite{MAGIC-EBL-1011} to analyze data from several 
AGNs observed with MAGIC. The aim is to investigate consistency of the results and to 
improve sensitivity to the EBL density at different redshifts.

\section{DATA SET}
The data set consists of 12 different sources, at redshift range from z=0.03 up to z=0.9,
taken with the MAGIC telescopes between 2010 and 2016.
The following criteria were applied to use AGN spectra in this analysis:
\begin{itemize}
 \item The data are taken with MAGIC in stereoscopic mode to reduce systematics effects. 
 \item The VHE spectra are selected, which are sensitive to the EBL density, i.e. strong signals, 
hard spectra and/or signals from distant sources. 
\end{itemize}
Apart from Mrk 421 (15 energy spectra) and PG\,1553+113 (5 spectra), there is only one averaged
spectrum per source. The criterion to split data set of one source to more than one spectrum
is based on source spectral shape variability at VHE $\gamma$ rays. 
In total, there are 30 independent energy spectra
used in this analysis. After quality cuts, the live time of the data set is 289.2\,h. 
The data set is summarized in Table~\ref{tab:dataset}.

\begin{table}[h]
\caption{The data set used in this analysis.}
\label{tab:dataset}
\tabcolsep7pt\begin{tabular}{lcccc}
\hline
\tch{1}{c}{b}{Source \\ Name}  & \tch{1}{c}{b}{AGN\\ Type}  & \tch{1}{c}{b}{Redshift \\ z}  
& \tch{1}{c}{b}{Data Taken \\ Period}  & \tch{1}{c}{b}{Effective \\ Time (h)}   \\
\hline
Mrk\,421 & HBL & 0.031 & 2013 Apr 10-19, 2014 Apr 26 & 40.4 \\
1ES\,1959+650 & HBL & 0.048 & 2015 Nov 6-18  & 4.8 \\
OT\,546 (1ES\,1727+502) & HBL & 0.055 & 2015 Oct 11 - Nov 2  & 6.4 \\
BL Lacertae & HBL & 0.069 & 2015 June 15  & 1.0 \\
1ES\,0229+200 & HBL & 0.140 & 2012 - 2015          & 105.2 \\
1ES\,1011+496 & HBL & 0.212 & 2014 Feb - Mar 7     & 11.8  \\
PKS\,1510-089 & FSRQ & 0.361 & 2015 May 18 - 19     & 2.4  \\
PKS\,1222+216 & FSRQ & 0.432 & 2010 Jun 18     & 0.5  \\
PG\,1553+113  & HBL  & 0.43-0.58 & 2012 - 2016 & 66.3  \\
PKS\,1424+240 & HBL  & 0.601 & 2014 Mar 24 - Jun 18   & 28.2  \\
PKS\,1441+25  & FSRQ & 0.939 & 2015 Apr 18 - 23   & 20.1  \\
B\,0218+35    & FSRQ & 0.944 & 2014 Jul 25 - 26   & 2.1  \\
{\bf {Total}}  &     &     &       &  {\bf {289.2}}   \\
\hline
\end{tabular}
\end{table}

\section{ANALYSIS}

\begin{figure}[h]
\centerline{\includegraphics[width=0.45\textwidth]{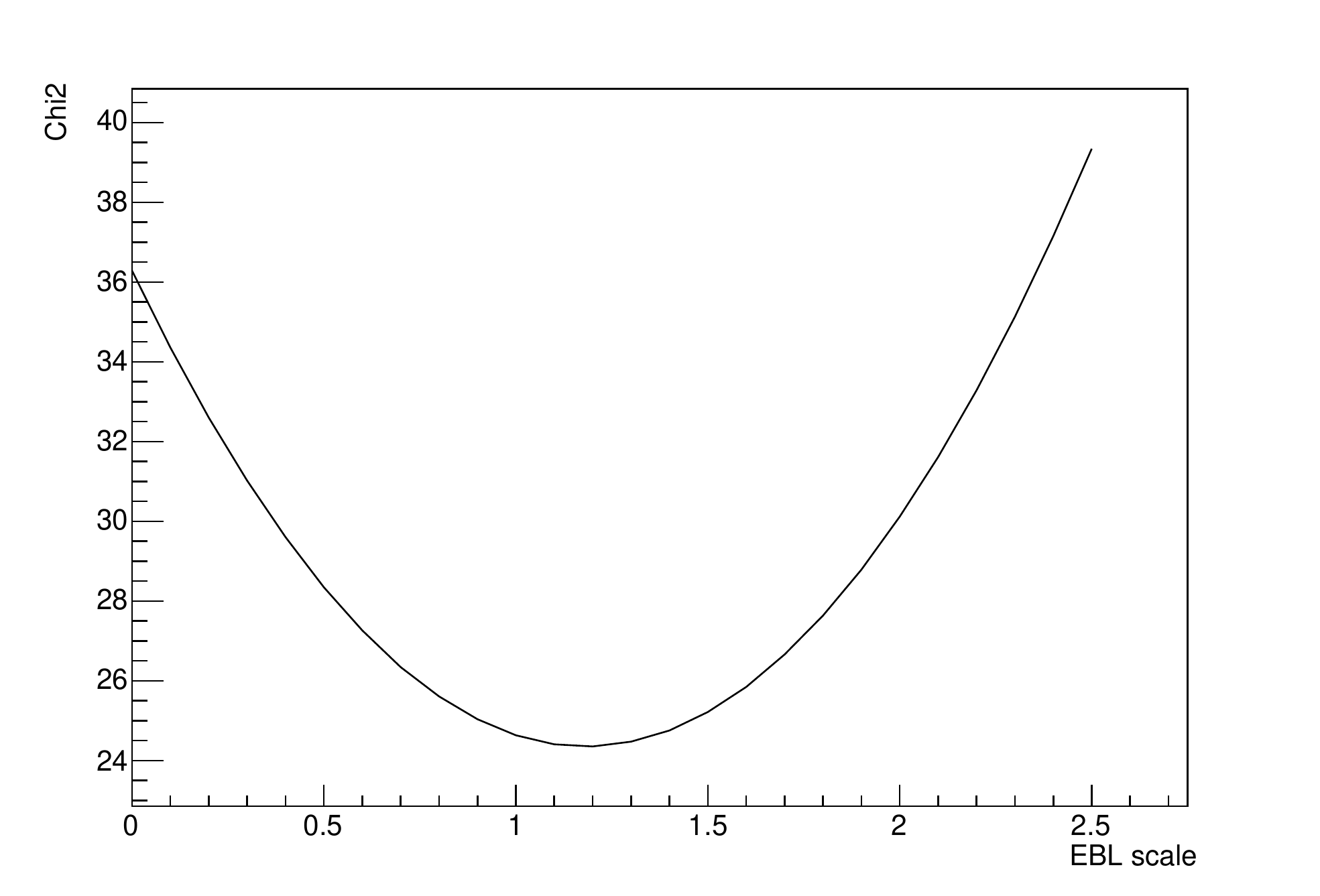}
\includegraphics[width=0.45\textwidth]{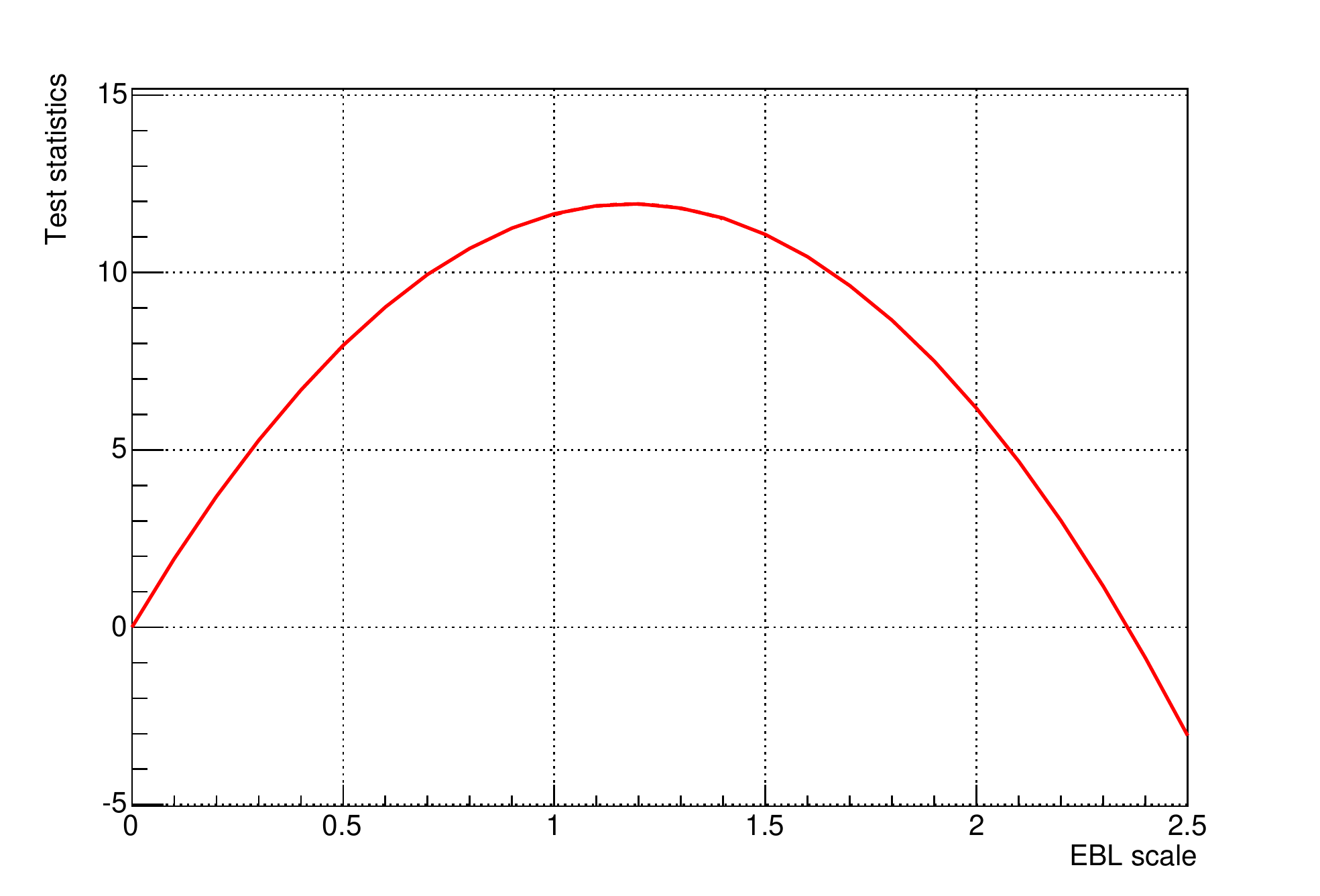}}
  \caption{Example of $\chi^2$ and TS distributions from a single energy spectrum as a function
of the EBL scale factor $\alpha$, see text for details.} 
\label{fig:example}
\end{figure}

We follow the procedure described in \cite{HESS2013} for the likelihood ratio test. The absorption of the EBL is described as $e^{-\alpha \tau(E,z)}$ where $\tau(E,z)$ is the optical depth predicted by the model, which depends on the energy $E$ of the $\gamma$-rays and the redshift $z$ of the source. With the optical depth scaled by a factor $\alpha$, the observed spectrum is formed as:
\begin{equation}
\left(\frac{dF}{dE}\right)_{obs}=\left(\frac{dF}{dE}\right)_{int} \times \exp(- \alpha \times \tau(E,z))
\end{equation}

\noindent where $(dF/dE)_{int}$ is the intrinsic spectrum of the source. The
emission of HBLs, like 1ES 1011+496, is often well described by basic
synchrotron self-Compton (SSC) models \citep[e.g.][]{Tavecchio1998}. A
population of electrons is accelerated to ultrarelativistic energies with a
resulting power-law spectrum with index $\Gamma_{e}$ of about 2. The high
energy electrons are cooled faster than the low energy ones, resulting in a
steeper $\Gamma_{e}$. These electrons produce synchrotron radiation with a
photon index $\Gamma=\frac{\Gamma_{e}+1}{2}=1.5$. In the Thomson regime the
energy spectrum index of the inverse-Compton scattered photons is approximately
the same as the synchrotron energy spectrum, whereas in the Klein-Nishima
regime, the resulting photon index is even larger. These arguments put serious
constraints to the photon index of the energy spectrum of VHE photons.
Additionally, in most of the SSC models, the emission is assumed to be
originated in a single compact region, which results in a smooth spectral
energy distribution with two concave peaks. The shape of the individual peaks
could be modified in a multizone model, where the emission is a superposition
of several one-zone emission regions. However the general two-peak structure is
conserved.

For the modeling of the intrinsic source spectrum we have used the same
functions as in \cite{Mazin2007} and \cite{HESS2013} which were also used to
fit the observed spectrum: power law (PWL), log-parabola (LP), 
power law with an exponential cut-off (EPWL), log-parabola with an exponential cut-off (ELP) 
and power law with a super or sub-exponential cut-off (SEPWL). We have added the
additional constraint that the shapes cannot be convex, i.e.\ the hardness of
the spectrum cannot increase with energy, as this is not expected in emission
models, nor has it been observed in any BL Lac in the optically-thin regime. In
particular, the un-absorbed part of BL Lac spectra measured by Fermi-LAT are
well fitted by log-parabolas \citep{Fermi2012}.

The analysis is performed in two steps.
First, for each energy spectrum and a given EBL energy density scaling factor $\alpha$
the Poisson likelihood function is calculated on the level of observed ON events in bins
of estimated energy while the spectral parameters of the 
intrinsic fit functions are left to vary freely in the optimization process. 
In addition the Poisson parameters of the background in each bin of
estimated energy are used as nuisance parameters in the likelihood maximization. 
We then make a scan over the EBL energy density scaling factor $\alpha$,
see an example of the resulting distribution in Figure~\ref{fig:example}, left plot.
The resulting $\chi^2$ values are converted into the test statistics (TS)
by taking the $\alpha=0$ (No-EBL) case as the null-hypothesis, see Figure~\ref{fig:example}, right plot.
In case of PG\,1553+113, as the redshift is uncertain, we treated the redshift
in the range z=(0.43-0.58) as an additional nuisance parameter.
In the second step, we linearly combine the TS distributions from the 30 individual spectra
and obtain the final result.

\subsection{Comments on fit function of intrinsic AGN spectra}

We would like to note that using PWL fit function can introduce a bias towards a too high
level of the EBL density, i.e.\ implying the EBL density is higher than it actually is.
The bias can be explained by the nature of the likelihood ratio test:
in case a function with more degrees of freedom than PWL is not preferred w.r.t. the PWL fit,
results of PWL would be used. The significance level for the preference is set to be 2$\sigma$.
This implies that an effect of 0-2$\sigma$ can be accumulated from individual sources each,
attributing all (not strongly pronounced per source) intrinsic curvature in the spectra to the EBL effect. 
We are, therefore, convinced that using a PWL fit to describe intrinsic spectra is not appropriate for setting EBL constraints. The simplest function (the one with the least number of free parameters) 
that we allow in our analysis is LP. 

The LP is a function that is linear in its parameters in the log
flux--log E representation (hence well-behaved in the fitting process), and
can model pretty well the de-absorbed spectra. 
The EPWL, ELP and SEPWL have additional (non-linear)
parameters that are physically motivated, e.g.\ to account for possible internal
absorption at the source. Note that these functions (except the PWL) can also
mimic the \textit{overall} spectral curvature induced by the EBL over a wide
range of redshifts, but will be unable to fit the inflection point (in the
optical depth vs. log E curvature) that state-of-the-art EBL models predict
around 1 TeV. We therefore expect an improvement of the fit quality as we
approach the true value of the scaling factor $\alpha$, hence providing a
constraint of the actual EBL density. The chosen spectral functions, however,
do not exhaust \textit{all possible} concave shapes. Therefore the EBL
constraints we will obtain are valid under the assumption that the true
intrinsic spectrum can be well described (within the uncertainties of the
recorded fluxes) by one of those functions.

\section{RESULTS}

\subsection{Statistical result}

\begin{figure}[h]
  \centerline{\includegraphics[width=0.75\textwidth]{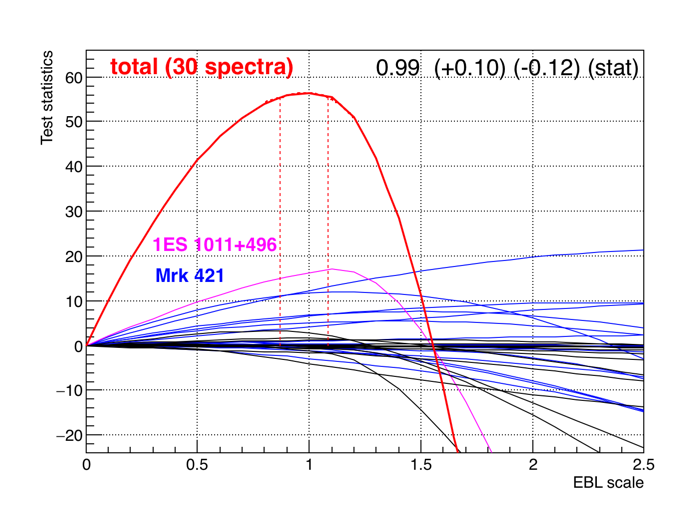}}
  \caption{The test statistics results of the analysis in this paper as a function of the 
EBL scaling factor. The thick red line represents the combined result of all spectra used,
the thin lines denote individual spectra. The dashed vertical lines represent 68\%
confidence level band of the combined result. The result obtained from the 1ES\,1011+496 spectrum 
alone is highlighted in magenta, the ones from Mrk\,421 are in blue. }
\label{fig:TS}
\end{figure}

The combined result is shown in Figure~\ref{fig:TS}.
The TS distributions of individual spectra are shown by thin black lines. 
To highlight the most constraining cases,
we plot the case of 1ES\,1011+496 in magenta and Mrk\,421 in blue.
The overall TS distribution is shown by the thick red line, and the 68\%
confidence level is marked by the vertical dashed lines.
We obtain $\alpha = 0.99 (+0.10) (-0.12)$ in this analysis, which is remarkably
close to the used EBL model by \cite{Dominguez2011}.
Here, only statistical errors are taken into account.

\subsection{Systematic uncertainty}

The MAGIC telescopes has a systematic uncertainty in the absolute energy scale
of 15\% \citep{MAGICperformance2014}. The main source of this uncertainty is
the imprecise knowledge of the atmospheric transmission. In order to assess how
this uncertainty affects the EBL constraint, the calibration constants used to
convert the pixel-wise digitized signals into photoelectrons were multiplied by
a scaling factor (the same for both telescopes) spanning the range between 0.85 and 1.15,
in steps of 0.05. This procedure is similar as the one presented by
\cite{MAGICperformance2014}.

For each of the scaling factors the data were processed in an identical manner through the full analysis chain, starting from the image cleaning, and using in all cases the standard MAGIC MC for this observation period. In this way we try to asses the effect of a potential miscalibration between the data and the MC simulation. 
The resulting EBL density using this procedure is
$\alpha = 0.99 (+0.15) (-0.56)$, where the errors is a combination
of the statistical and systematic uncertainties.

\subsection{Comment on possible upturn in energy spectra at high $\tau$ values}

\begin{figure}[h]
  \centerline{\includegraphics[width=0.95\textwidth]{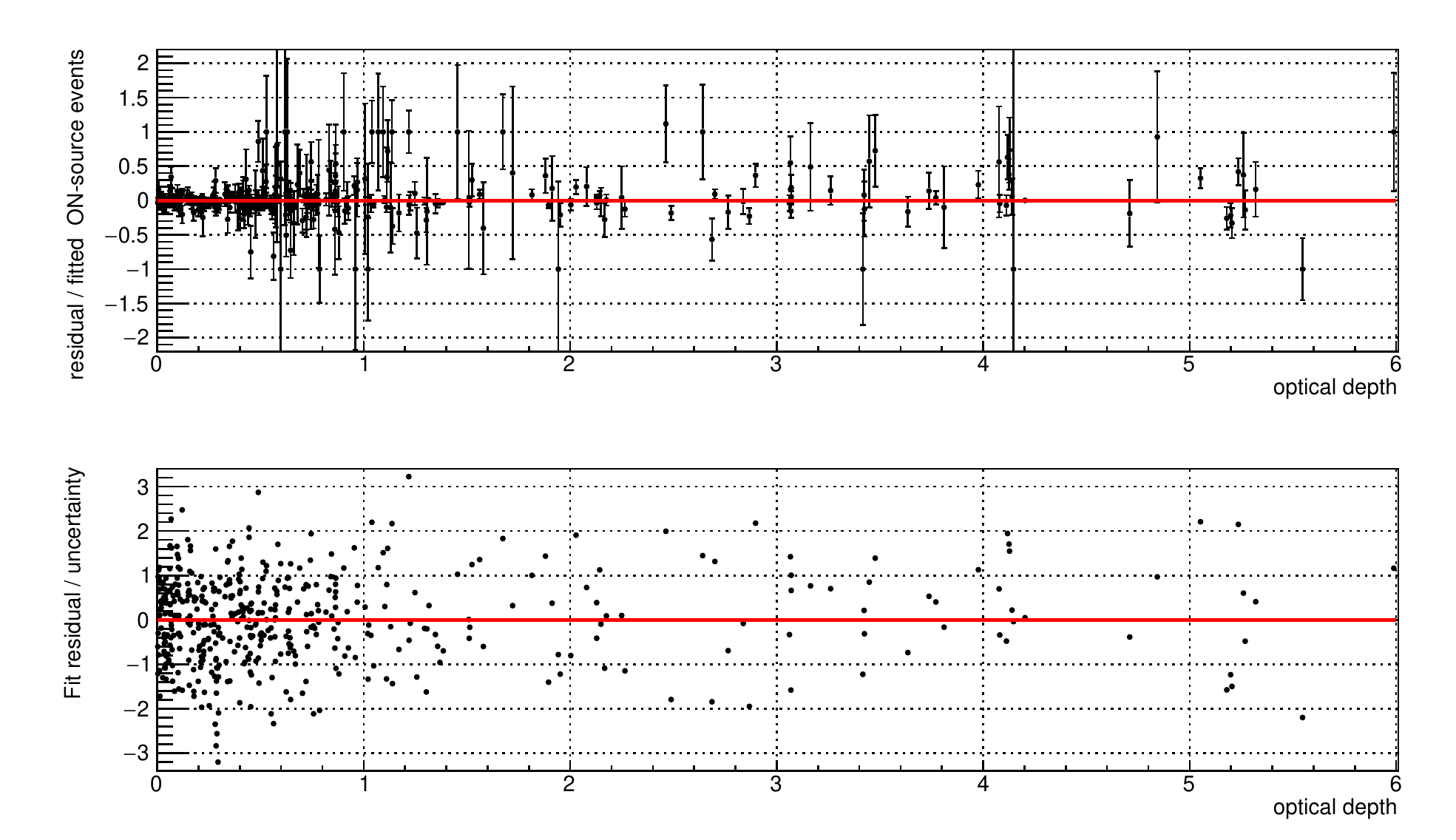}}
  \caption{The residuals between ON source events and best spectral fit function as a function
of the optical depth. No cuts are applied to the significance of the individual points.
Upper panel shows the residuals and the corresponding error bars. Lower panel shows uncertainties
of the residuals. There is no clear trend observed.}
\label{fig:nopileup}
\end{figure}

There have been claims that the residuals between measured energy spectra of extragalactic sources
and the best-fit models start to systematically deviate towards higher values of the optical depth $\tau$
\citep[see, e.g.,][]{Horns2012a}. We investigated this on our sample.  When no
cuts are applied (e.g., on minimum statistical significance) to the individual
reconstructed spectral points in our sample, we did not find any trend towards
higher $\tau$-values, see Figure~\ref{fig:nopileup}.  However, there is some
trend visible, though not as strong as in \cite{Horns2012a}, when we remove
non-significant points. We, therefore, suggest that the trend found in previous
studies is at least partly due to an observational bias, if spectral points 
only above a certain minimum significance are included in publications. Such procedure,
especially in case of soft spectra, introduces some bias towards higher energy
end of the spectra (thus, towards higher optical depths) where the event
statistics is low and only upward fluctuations survive the cuts.

\section{CONCLUSION}

In this paper we showed results of combined analysis of 30 energy spectra from
11 sources in the redshift range from z=0.03 up to z=0.9.
The method used is similar to \cite{Fermi2012, HESS2013} and \cite{MAGIC-EBL-1011}.
The EBL model used here is from \cite{Dominguez2011} and similar results
are obtained when using models of \cite{Franceschini2008,Finke2010} and \cite{Gilmore2012}.
Our result on the EBL scaling factor is $\alpha = 0.99 (+0.10) (-0.12)$,
statistical errors only and 
$\alpha = 0.99 (+0.15) (-0.56)$ when adding systematic uncertainties.
We will further investigate the relatively high systematic uncertainty on the lower side of the
measurement. In particular, we note that the uncertainty in the absolute energy scale assumed here
to be 15\%
(dominating our systematic uncertainty) is valid for individual
spectra taken in a short exposure time.  For a set of spectra taken at
different years, at different weather conditions, and different zenith angles
the resulting average systematic uncertainty in the energy scale is significantly
smaller.  Therefore, our error estimation should be considered as
conservative.

We also investigated a possible upturn or trend in the residuals between the
best fit functions and the measured excess events versus the optical depth.
However, no trend is found in our data set and we suggest that a positive bias
in previously published energy spectra for points with low statistics 
may be causing the observed trend.

\section{ACKNOWLEDGMENTS}
We would like to thank
the Instituto de Astrof\'{\i}sica de Canarias
for the excellent working conditions
at the Observatorio del Roque de los Muchachos in La Palma.
The financial support of the German BMBF and MPG,
the Italian INFN and INAF,
the Swiss National Fund SNF,
the he ERDF under the Spanish MINECO
(FPA2015-69818-P, FPA2012-36668, FPA2015-68278-P,
FPA2015-69210-C6-2-R, FPA2015-69210-C6-4-R,
FPA2015-69210-C6-6-R, AYA2013-47447-C3-1-P,
AYA2015-71042-P, ESP2015-71662-C2-2-P, CSD2009-00064),
and the Japanese JSPS and MEXT
is gratefully acknowledged.
This work was also supported
by the Spanish Centro de Excelencia ``Severo Ochoa''
SEV-2012-0234 and SEV-2015-0548,
and Unidad de Excelencia ``Mar\'{\i}a de Maeztu'' MDM-2014-0369,
by grant 268740 of the Academy of Finland,
by the Croatian Science Foundation (HrZZ) Project 09/176
and the University of Rijeka Project 13.12.1.3.02,
by the DFG Collaborative Research Centers SFB823/C4 and SFB876/C3,
and by the Polish MNiSzW grant 745/N-HESS-MAGIC/2010/0.


\nocite{*}
\bibliographystyle{ieeetr}
\bibliography{eblmagic}%

\end{document}